\documentclass[
aps,prb,reprint,superscriptaddress,showpacs]{revtex4-2}

\usepackage{graphicx}
\usepackage{bm}
\usepackage{amsmath}
\usepackage{amsfonts}
\usepackage{amssymb}
\usepackage{braket}
\usepackage{dsfont}
\usepackage{eucal}
\usepackage{mathtools}
\usepackage{comment}
\usepackage{bbm}
\usepackage{tikz}
\usepackage{tikz-3dplot}
\usetikzlibrary{calc,arrows,backgrounds,shapes,mindmap,decorations.markings,plotmarks}
\usepackage{array}
\usepackage{parskip}
\usepackage{hyperref}
\usepackage{amsmath,amssymb}
\usepackage{calrsfs}

\definecolor{myolive}{RGB}{181,204,194}

\definecolor{myred}{RGB}{204,102,102}
\definecolor{myblue}{RGB}{120,180,190}
\definecolor{myyellow}{RGB}{230,210,140} 
\definecolor{mypurple}{RGB}{170,140,190} 
\definecolor{mygreen}{RGB}{140,190,150}   
\definecolor{myorange}{RGB}{220,160,100}   
\newcommand{\nn}{\nonumber}

\def\ri{{\rm i}}

\newcommand{\trtenH}[4]{
\draw[thick] (#1,#2+0.25)--(#1-0.1,#2+0.25);
\draw[thick] (#1-0.1,#2+0.25) to[out=180,in=180] (#1-0.1,#2-0.1);
\draw[thick] (#3+0.5,#4+0.25)--(#3+0.5+0.1,#4+0.25);
\draw[thick] (#3+0.5+0.1,#4+0.25) to[out=0,in=0] (#3+0.5+0.1,#4-0.1);
\draw[thick] (#3+0.5+0.1,#4-0.1) -- (#1-0.1,#2-0.1);
}

\newcommand{\tightldots}{\mathinner{\ldotp\mkern-2mu\ldotp\mkern-2mu\ldotp}}

\newcommand{\MPSxOnexD}[8]{
\filldraw[fill=myolive,thick, rounded corners=4pt] (#2+0,#3+0) rectangle (#2+#4,#3+#5);
\draw[thick] (#2,#3+0.5*#5)--(#2-#6,#3+0.5*#5);
\draw[thick] (#2+#4,#3+0.5*#5)--(#2+#4+#7,#3+0.5*#5);
\draw[thick] (#2+0.5*#4,#3+#5)--(#2+0.5*#4,#3+#5+#8);

\node[] at (#2+0.5*#4,#3+0.5*#5) {\scalebox{0.8}{$#1$}};}

\newcommand{\MPSxOnexDBlue}[8]{
\filldraw[fill=myblue,thick, rounded corners=4pt] (#2+0,#3+0) rectangle (#2+#4,#3+#5);
\draw[thick] (#2,#3+0.5*#5)--(#2-#6,#3+0.5*#5);
\draw[thick] (#2+#4,#3+0.5*#5)--(#2+#4+#7,#3+0.5*#5);
\draw[thick] (#2+0.5*#4,#3+#5)--(#2+0.5*#4,#3+#5+#8);

\node[] at (#2+0.5*#4,#3+0.5*#5) {\scalebox{0.8}{$#1$}};}

\newcommand{\MPSxOnexDColor}[9]{
\filldraw[fill=#9,thick, rounded corners=4pt] (#2+0,#3+0) rectangle (#2+#4,#3+#5);
\draw[thick] (#2,#3+0.5*#5)--(#2-#6,#3+0.5*#5);
\draw[thick] (#2+#4,#3+0.5*#5)--(#2+#4+#7,#3+0.5*#5);
\draw[thick] (#2+0.5*#4,#3+#5)--(#2+0.5*#4,#3+#5+#8);

\node[] at (#2+0.5*#4,#3+0.5*#5) {\scalebox{0.8}{$#1$}};}

\newcommand{\drawLegs}[8]{%
  \pgfmathsetmacro{\cx}{#1}
    \pgfmathsetmacro{\cy}{#2}
    \pgfmathsetmacro{\R}{#3}
    \pgfmathsetmacro{\start}{#4}
    \pgfmathsetmacro{\endang}{#5}
    \pgfmathsetmacro{\step}{#6}
    \pgfmathtruncatemacro{\Nlegs}{#7}

    \pgfmathsetmacro{\tol}{0.01}

    \coordinate (C) at (\cx,\cy);

    \pgfmathsetmacro{\legsep}{(\endang-\start)/(\Nlegs+1)}

    \foreach \k in {1,...,\Nlegs} {
      \pgfmathsetmacro{\legangle}{\start + \k*\legsep}
      \ifodd\k
  \draw[
    thick,
    postaction={
      decorate,
      decoration={
        markings,
        mark=at position 0.90 with {\arrow{>}}
      }
    }
  ] (C) -- ($(C)+(\legangle:\R)$);
\else
  \draw[
    thick,
    postaction={
      decorate,
      decoration={
        markings,
        mark=at position 0.90 with {\arrow{<}}
      }
    }
  ] (C) -- ($(C)+(\legangle:\R)$);
\fi
      
    }

    \pgfmathtruncatemacro{\Ndots}{floor((\endang-\start)/\step)-1}

    \ifnum\Ndots>0
      \foreach \i in {1,...,\Ndots} {
        \pgfmathsetmacro{\ang}{\start + \i*\step}
        \pgfmathsetmacro{\drawdot}{1}

        \foreach \k in {1,...,\Nlegs} {
          \pgfmathsetmacro{\legangle}{\start + \k*\legsep}
          \pgfmathparse{abs(\ang-\legangle)<\tol ? 0 : \drawdot}
          \let\drawdot\pgfmathresult
        }

        \ifnum\drawdot=1
          \fill ($(C)+(\ang:{#8*\R})$) circle (0.4pt);
        \fi
      }
    \fi

}

\newcommand{\PEPS}[6]{
\def\x{#1}
\def\y{#2}
\def\z{#3}
\def\width{#4}
\def\legSizeX{#5}
\def\legSizeY{#6}

\draw[thick, rounded corners=1pt, fill=myolive]
  (\x,\y,\z) -- (\x+\width,\y,\z) -- (\x+\width,\y,\z+\width) -- (\x,\y,\z+\width) -- cycle;
\draw[thick] (\x+0.5*\width,\y,\z+0.5*\width) -- (\x+0.5*\width,\y+0.5*\width,\z+0.5*\width);
\draw[thick] (\x+0.5*\width,\y,\z+0.5*\width) -- (\x+0.5*\width,\y+0.5*\width,\z+0.5*\width);
\draw[thick] (\x+0.5*\width,\y,\z+0.5*\width) -- (\x+0.5*\width,\y+0.5*\width,\z+0.5*\width);

\draw[thick] (\x,\y,\z+0.5*\width)--(\x-\legSizeX,\y,\z+0.5*\width);
\draw[thick] (\x+\width,\y,\z+0.5*\width)--(\x+\width+\legSizeX,\y,\z+0.5*\width);
\draw[thick] (\x+0.5*\width,\y,\z+\width)--(\x+0.5*\width,\y,\z+\width+\legSizeY);
\draw[thick] (\x+0.5*\width,\y,\z)--(\x+0.5*\width,\y,\z-\legSizeY);
}

\newcommand{\PEPSY}[7]{
\def\x{#1}
\def\y{#2}
\def\z{#3}
\def\width{#4}
\def\legSizeX{#5}
\def\legSizeY{#6}
\def\legSizeZ{#7}

\draw[thick, rounded corners=1pt, fill=myolive]
  (\x,\y,\z) -- (\x+\width,\y,\z) -- (\x+\width,\y,\z+\width) -- (\x,\y,\z+\width) -- cycle;
\draw[thick] (\x+0.5*\width,\y,\z+0.5*\width) -- (\x+0.5*\width,\y+\legSizeZ,\z+0.5*\width);
\draw[thick] (\x+0.5*\width,\y,\z+0.5*\width) -- (\x+0.5*\width,\y+0.5*\width,\z+0.5*\width);
\draw[thick] (\x+0.5*\width,\y,\z+0.5*\width) -- (\x+0.5*\width,\y+0.5*\width,\z+0.5*\width);

\draw[thick] (\x,\y,\z+0.5*\width)--(\x-\legSizeX,\y,\z+0.5*\width);
\draw[thick] (\x+\width,\y,\z+0.5*\width)--(\x+\width+\legSizeX,\y,\z+0.5*\width);
\draw[thick] (\x+0.5*\width,\y,\z+\width)--(\x+0.5*\width,\y,\z+\width+\legSizeY);
\draw[thick] (\x+0.5*\width,\y,\z)--(\x+0.5*\width,\y,\z-\legSizeY);
}

\newcommand{\PEPSC}[7]{
\def\x{#1}
\def\y{#2}
\def\z{#3}
\def\width{#4}
\def\legSizeX{#5}
\def\legSizeY{#6}

\draw[thick, rounded corners=1pt, fill=#7]
  (\x,\y,\z) -- (\x+\width,\y,\z) -- (\x+\width,\y,\z+\width) -- (\x,\y,\z+\width) -- cycle;
\draw[thick] (\x+0.5*\width,\y,\z+0.5*\width) -- (\x+0.5*\width,\y+0.5*\width,\z+0.5*\width);
\draw[thick] (\x+0.5*\width,\y,\z+0.5*\width) -- (\x+0.5*\width,\y+0.5*\width,\z+0.5*\width);
\draw[thick] (\x+0.5*\width,\y,\z+0.5*\width) -- (\x+0.5*\width,\y+0.5*\width,\z+0.5*\width);

\draw[thick] (\x,\y,\z+0.5*\width)--(\x-\legSizeX,\y,\z+0.5*\width);
\draw[thick] (\x+\width,\y,\z+0.5*\width)--(\x+\width+\legSizeX,\y,\z+0.5*\width);
\draw[thick] (\x+0.5*\width,\y,\z+\width)--(\x+0.5*\width,\y,\z+\width+\legSizeY);
\draw[thick] (\x+0.5*\width,\y,\z)--(\x+0.5*\width,\y,\z-\legSizeY);
}

\newcommand{\be}{\begin{equation}}
\newcommand{\ee}{\end{equation}}
\newcommand{\bea}{\begin{eqnarray}}
\newcommand{\eea}{\end{eqnarray}}

\def\nn{\nonumber\\}

\begin{document}

\title{Exact Quantum Many-Body Scars by a generalized Matrix-Product Ansatz}

\author{Sascha Gehrmann and Fabian H.L. Essler}
\affiliation{The Rudolf Peierls Centre for Theoretical Physics,
Oxford University, Oxford OX1 3PU, UK}

\begin{abstract}
We construct exact eigenstates of quantum many-body systems with Hamiltonians that are not frustration-free in matrix product form, based on a local error cancellation ansatz motivated by the Derrida-Evans-Hakim-Pasquier method for finding the stationary state of the asymmetric simple exclusion process. We demonstrate the approach with explicit examples in both one and two spatial dimensions.
\end{abstract}

\maketitle
%%%%%%%%%%%%%%%%%%%%%%%%%
\section{Introduction}
%%%%%%%%%%%%%%%%%%%%%%%%%%
The search for ergodicity-breaking mechanisms in the non-equilibrium dynamics of interacting many particle quantum systems has attracted an enormous amount attention in recent years. Known ways of completely or partially circumventing ergodic behavior include Yang-Baxter integrability \cite{essler2016quench}, sufficiently strong quenched disorder \cite{abanin2019colloquium}, kinematic constraints \cite{pancotti2020quantum}, Hilbert space fragmentation \cite{moudgalya2022quantum} and quantum many-body scars \cite{serbyn2021quantum,chandran2023quantum}. All these mechanisms are related to the existence of energy eigenstates at finite energy densities above the ground state whose local properties are not thermal. Among these, quantum many-body scars have been of particular interest because they can have low entanglement and allow for relatively simple, closed form expressions \cite{ShiMor2017}. This in turn can make them experimentally accessible in ultracold atom quantum simulators \cite{Bernien2017,Turner2018a,Turner2018b,Jepsen2020,Jepsen2022}
and quantum processors \cite{Zhang2023}. Over the last few years, several frameworks for constructing scar eigenstates
at finite energy densities have been developed. These include
embedding constructions \cite{Shiraishi2019,Omiya2023}, group-theoretic structures for degenerate multiplets \cite{Bull2020,Pakrouski2020,ODea2020,Tang2022,Pakrouski2021,Ren2021}, spin-helix states, \cite{Lange:2025svs,Wang:2024tso,Hu:2025zvv,Bhowmick:2025dip,PhysRevLett.132.220404,PhysRevB.106.075406,PhysRevB.111.094437} 
and spectrum-generating algebras that provide raising and lowering operators to build towers of eigenstates with equally spaced energies \cite{Moudgalya_2018_exact,Moudgalya_2018,Iadecola2019,Schecter2019,Moudgalya_2020,Moudgalya_2020_eta,Chattopadhyay2020,Iadecola2020,mark2020unified,mark2020eta,Chandran2023}.
Scars have also been constructed in Floquet systems \cite{Mukherjee2019,Pai2019,Haldar2021,Schmid2026}, random unitary circuits \cite{Capizzi2026}, disordered systems exhibiting Onsager-like structures \cite{Shibata2019}.

The scar eigenstates constructed in the literature are mostly simple by design. In particular they can take the form of matrix product states (MPS). One well-established way of obtaining exact eigenstates $|\Psi\rangle$ of Hamiltonians of the form
\be
H=\sum_{\langle j,k\rangle} h_{j,k}\ ,
\ee
where the sum is over links $\langle j,k\rangle$ on a D-dimensional lattice, is by imposing a \emph{frustration-free}
condition
\be
h_{j,k}|\Psi\rangle=0\quad \forall \text{ links }\langle{j,k}\rangle.
\label{frustfree}
\ee
This has been employed to great effect to obtain exact MPS ground states of spin models
\cite{Affleck:1987,Fannes1992,PerezGarcia2007,KluSchZit1993,NigKluZit1997,NigKluZit2000,Schuch:2010}, and more recently to construct exact quantum many-body scars \cite{PhysRevLett.131.020402,PhysRevLett.122.173401,Lee2020,xu2025extracting}.

It is then very natural to ask whether it is possible to construct algorithms for obtaining exact MPS eigenstates that relax the frustration-free condition (\ref{frustfree}). A number of works have addressed this question
using a range of approaches \cite{Moudgalya_2018_exact,PhysRevLett.122.173401,Moudgalya_2020,ODea2020,PhysRevLett.122.173401,Katsura,matsui2024exactly,moudgalya2024exhaustive,ivanov2025many,gioia2025distinct}. In particular, Refs
\cite{Katsura,matsui2024exactly} considered examples of a more general method for constructing exact matrix-product eigenstates in spin models without frustration-free structures presented in Section~\ref{sec:DEHP}. This method is a generalization of the Derrida-Evans-Hakim-Pasquier (DHEP) ansatz \cite{DerEvaHakPas1993} for the exact steady state of the asymmetric simple exclusion process with open boundaries, and in particular its reduction to finite bond dimensions \cite{EssRit1996,mallick1997finite}. 
%%%%%%%%%%%%%%%%%%%%%%%%%%%%%%%%%%%%%%
\section{Generalized DEHP Ansatz} 
\label{sec:DEHP}
%%%%%%%%%%%%%%%%%%%%%%%%%%%%%%%%%%%%%%
Let $\mathcal{G} = (\mathcal{V}, \mathcal{E})$ be general oriented graph, where $\mathcal{V}$ and $\mathcal{E} \subset \mathcal{V} \times \mathcal{V}$ denote respectively the sets of $N_v$ nodes and $N_e$ oriented edges. For each node $v \in \mathcal{V}$, we further define the set $\mathcal{E}_v$ of edges connected to $v$.  We now consider a quantum spin system obtained by placing a qudit of dimension $d$ on each node of the graph, and defining a local Hamiltonian with nearest-neighbour interactions of the form
\begin{align}
    H=\sum_{(i,j)\in \mathcal{E}} \, h_{i,j}\,.
\end{align}
Our objective is to construct exact eigenstates $\ket{\Psi}$ of $H$ of Matrix Product State (MPS) form
\begin{align}
    \ket{\Psi}=\sum_{\{s_v| v\in \mathcal{V}\}} \sum_{\{\alpha_{e}|e\in \mathcal{E}\}} 
\raisebox{-0.3em}{\Big(} \prod_{v \in \mathcal{V}} [A_v]^{s_v}_{ \{\alpha_{k}|k \in \mathcal{E}_v\}} \raisebox{-0.3em}{\Big) }
|\{s_v\}\rangle\ .
\label{MPSform}
\end{align}
Here $\{|\{s_v\}\rangle| v\in \mathcal{V}\}$ is a basis of product states, where $s_v\in\{1,2,\dots,d\}$ labels the $d$ eigenstates of a single qudit and
\begin{align}
   [A_v]^{s_v}_{\{\alpha_e|e\in \mathcal{E}_v\}}
\end{align}
are tensors with one physical index $s_v$ of dimension $d$, and a set of auxiliary indices $\alpha_e\in\{1,\dots,\chi\}$.
To ease notations, we will henceforth suppress free indices and sum over repeated indices. The $A$-tensors can be represented diagrammatically as
\begin{center}
\begin{tikzpicture}
\drawLegs{0.25}{0.25}{0.65}{180}{360}{15}{2}{0.75}
\draw[thick,red] (0.25,0.5)--(0.25,0.75);
\draw[thick,->] (0,0.25)--(-0.25,0.25);
\draw[thick,-] (0,0.25)--(-0.35,0.25);
\draw[thick,-<] (0.5,0.25)--(0.5+0.25,0.25);
\draw[thick,-] (0.5,0.25)--(0.5+0.35,0.25);
\MPSxOnexD{A_v}{0}{0}{0.5}{0.5}{0.25}{0.25}{0}
\node[right] at (-3,0.25) {$[A_v]^{\textcolor{red}{s_v}}_{\{\alpha_{e}|e\in\mathcal{E}_v\}}\,\,=$};
\end{tikzpicture}  
\end{center}
We propose the following generalization of a frustration-free condition for constructing eigenstates of $H$: for each pair of nodes $k$ and $\ell$ connected by an oriented edge pointing from $k$ to $\ell$ we impose
\begin{align}\label{gnkdnrnr}
 &h_{k,\ell}\, [A_k]^{s_k}_{\alpha_{(k,\ell)}} [A_\ell]^{s_\ell}_{\alpha_{(k,\ell)}}\ket{s_k,s_\ell}\\&= \big( [E^{(k,l)}_k]^{s_k}_{\alpha_{(k,\ell)}}[A_\ell]^{s_\ell}_{\alpha_{(k,\ell)}}+[A_k]^{s_k}_{\alpha_{(k,\ell)}} [E^{(k,\ell)}_\ell]^{s_\ell}_{\alpha_{(k,\ell)}}\big)\ket{s_k,s_\ell}\notag
\end{align}
where $E^{(k,\ell)}_{k}$,$E^{(k,\ell)}_{\ell}$  are  tensors of the same kind as $A$. In the following, we refer to the right-hand-side of eqn (\ref{gnkdnrnr}) as error terms. In terms of diagrams the equation reads
\begin{center}
\begin{tikzpicture}

\pgfmathsetmacro{\dowOffSet}{-0.35}
\drawLegs{0.25}{0.25+\dowOffSet}{0.65}{180}{360}{15}{2}{0.75}
\draw[thick,red] (0.25,0.5+\dowOffSet)--(0.25,0.75+\dowOffSet);
\draw[thick,->] (0,0.25+\dowOffSet)--(-0.25,0.25+\dowOffSet);
\draw[thick,-] (0,0.25+\dowOffSet)--(-0.35,0.25+\dowOffSet);
\draw[thick,-<] (0.5,0.25+\dowOffSet)--(0.5+0.25,0.25+\dowOffSet);
\draw[thick,-] (0.5,0.25+\dowOffSet)--(0.5+0.35,0.25+\dowOffSet);
\MPSxOnexD{A_k}{0}{0+\dowOffSet}{0.5}{0.5}{0.25}{0.25}{0}

\drawLegs{0.25+1}{0.25+\dowOffSet}{0.65}{180}{360}{15}{2}{0.75}
\draw[thick,red] (0.25+1,0.5+\dowOffSet)--(0.25+1,0.75+\dowOffSet);

\draw[thick,-] (0+1,0.25+\dowOffSet)--(-0.35+1,0.25+\dowOffSet);
\draw[thick,-<] (0.5+1,0.25+\dowOffSet)--(0.5+0.25+1,0.25+\dowOffSet);
\draw[thick,-] (0.5+1,0.25+\dowOffSet)--(0.5+0.35+1,0.25+\dowOffSet);
\MPSxOnexD{A_\ell}{+1}{0+\dowOffSet}{0.5}{0.5}{0.25}{0.25}{0}

\MPSxOnexDColor{h_{k,\ell}}{0}{0.75+\dowOffSet}{1.5}{0.5}{0}{0}{0}{myred}
\draw[thick,red] (0.25,0.5+0.75+\dowOffSet)--(0.25,0.75+0.75+\dowOffSet);
\draw[thick,red] (0.25+1,0.5+0.75+\dowOffSet)--(0.25+1,0.75+0.75+\dowOffSet);

\node[right] at (2,0.24) {$=$};

\drawLegs{0.25+3}{0.25}{0.65}{180}{360}{15}{2}{0.75}
\draw[thick,red] (0.25+3,0.5)--(0.25+3,0.75);
\draw[thick,-] (0+3,0.25)--(-0.35+3,0.25);
\draw[thick,->] (0+3,0.25)--(-0.25+3,0.25);
\draw[thick,-<] (0.5+3,0.25)--(0.5+0.25+3,0.25);
\draw[thick,-] (0.5+3,0.25)--(0.5+0.35+3,0.25);
\MPSxOnexDBlue{\scalebox{0.9}{$E^{k,\ell}_k$}}{+3}{0}{0.55}{0.5}{0.25}{0.25}{0}

\drawLegs{0.25+4}{0.25}{0.65}{180}{360}{15}{2}{0.75}
\draw[thick,red] (0.25+4,0.5)--(0.25+4,0.75);
\draw[thick,-] (0+4,0.25)--(-0.35+4,0.25);
\draw[thick,-<] (0.5+4,0.25)--(0.5+0.25+4,0.25);
\draw[thick,-] (0.5+4,0.25)--(0.5+0.35+4,0.25);
\MPSxOnexD{A_\ell}{+4}{0}{0.5}{0.5}{0.25}{0.25}{0}

\node[right] at (5,0.25) {$+$};

\pgfmathsetmacro{\x}{6}
\drawLegs{0.25+\x}{0.25}{0.65}{180}{360}{15}{2}{0.75}
\draw[thick,red] (0.25+\x,0.5)--(0.25+\x,0.75);
\draw[thick,->] (0+\x,0.25)--(-0.25+\x,0.25);
\draw[thick,-] (0+\x,0.25)--(-0.35+\x,0.25);
\draw[thick,-<] (0.5+\x,0.25)--(0.5+0.25+\x,0.25);
\draw[thick,-] (0.5+\x,0.25)--(0.5+0.35+\x,0.25);
\MPSxOnexD{A_k}{\x}{0}{0.5}{0.5}{0.25}{0.25}{0}

\pgfmathsetmacro{\x}{7}
\drawLegs{0.25+\x}{0.25}{0.65}{180}{360}{15}{2}{0.75}
\draw[thick,red] (0.25+\x,0.5)--(0.25+\x,0.75);
\draw[thick,-] (0+\x,0.25)--(-0.35+\x,0.25);
\draw[thick,-<] (0.5+\x,0.25)--(0.5+0.25+\x,0.25);
\draw[thick,-] (0.5+\x,0.25)--(0.5+0.35+\x,0.25);
\MPSxOnexDBlue{\scalebox{0.9}{$E^{{k,\ell}}_\ell$}}{\x}{0}{0.55}{0.5}{0.25}{0.25}{0}
        
    \end{tikzpicture}    
\end{center} 
where the blue and red boxes represent the $E$ tensors and the Hamiltonian density $h$.

Let us assume there exists a set of tensors $A$ and $E$'s that fulfil \eqref{gnkdnrnr}. If we require
\begin{align}
\sum_{e_v\in \mathcal{E}_v }E^{e_v}_v=0\qquad \forall \,v \in \mathcal{V}
\end{align}
then $|\Psi\rangle$ is an exact energy eigenstate with zero energy
\begin{align}\label{fanjnjn}
H|\Psi\rangle=0.
\end{align}
This follows from (\ref{gnkdnrnr}) once we sum over all links as the terms on the right-hand-side add up to zero.
%We note that analogous "telescoping" cancellations have previously appeared in other approaches to QMBS, see e.g.
%\cite{Moudgalya_2020,mark2020unified,mark2020eta}.
%%%%%%%%%%%%%%%%%%%%%%%%%%%%
\section{The method in 1D}
%%%%%%%%%%%%%%%%%%%%%%%%%%%%
We start by considering one-dimensional spin-S chains on a ring with $N$ sites. The local Hilbert space dimension is $d = 2S + 1$ and the Hamiltonian is taken to involve only interactions on neighbouring sites
\begin{align}
H=\sum^N_{j=1} \, h_{j,j+1}\,,
\label{Hgeneral}
\end{align}
where for the time being we impose periodic boundary conditions. We seek an exact eigenstate $\ket{\Psi}$ of $H$ in MPS form 
%\begin{align}
$\ket{\Psi}={\rm Tr}(AA\dots A)$
%\end{align}
where $A$ are  $\chi\times \chi$ matrices with $\mathbb{C}^d$-valued entries. In the standard graphical representation for MPS tensors \cite{penrose1971negative} we have 
\begin{center}
\begin{tikzpicture}
\draw[thick,red] (0,0.5)--(0,0.65);
%\draw[thick,->] (0,0.25)--(-0.25,0.25);
%\draw[thick,-] (0,0.25)--(-0.35,0.25);
%\draw[thick,-<] (0.5,0.25)--(0.5+0.25,0.25);
%\draw[thick,-] (0.5,0.25)--(0.5+0.35,0.25);
\MPSxOnexD{A}{-0.25}{0}{0.5}{0.5}{0.15}{0.15}{0}
\node[left] at (-0.3,0.25) {$\alpha$};
\node[right] at (0.3,0.25) {$\beta\,,$};
\node[above] at (0,0.6) {$\textcolor{red}{s}$};
\node[right] at (-2,0.2) {$A^{\textcolor{red}{s}}_{\alpha \beta}\,=$};

\node[right] at (0.9,0.25) {$\ket{\Psi}=$};
\draw[thick,red] (3.25-0.75,0.5)--(3.25-0.75,0.75);
\MPSxOnexD{A}{3-0.75}{0}{0.5}{0.5}{0.1}{0.15}{0}
\draw[thick,red] (3.25,0.5)--(3.25,0.75);
\MPSxOnexD{A}{3}{0}{0.5}{0.5}{0.25}{0}{0}

\draw[thick, dashed] (3.5,0.25)--(4.5,0.25);

\draw[thick,red] (4.75,0.5)--(4.75,0.75);
\MPSxOnexD{A}{4.5}{0}{0.5}{0.5}{0}{0.1}{0}

\trtenH{2.25}{0}{4.5}{0}
\end{tikzpicture}  
\end{center}
where the bond indices $\alpha,\beta$ run from $1$ to $\chi$ and the physical index $s$ from 1 to $d$. 
The condition (\ref{gnkdnrnr}) on the tensors $A$ takes the simple form
\begin{align}\label{dksndjsnfjdsnfjdnfjdnfj}
h\, A\, A =EA-AE\ ,
\end{align}
where $h$ is the Hamiltonian density and $E$ is a tensor of the same structure as $A$. Graphically eqn (\ref{dksndjsnfjdsnfjdnfjdnfj}) is represented as
\begin{equation*}
\begin{aligned}
\begin{tikzpicture}
\pgfmathsetmacro{\dowOffSet}{-0.35}
\draw[thick,red] (0.25,0.5+\dowOffSet)--(0.25,0.75+\dowOffSet);
%\draw[thick,->] (0,0.25+\dowOffSet)--(-0.25,0.25+\dowOffSet);
\draw[thick,-] (0,0.25+\dowOffSet)--(-0.35,0.25+\dowOffSet);
%\draw[thick,-<] (0.5,0.25+\dowOffSet)--(0.5+0.25,0.25+\dowOffSet);
\draw[thick,-] (0.5,0.25+\dowOffSet)--(0.5+0.35,0.25+\dowOffSet);
\MPSxOnexD{A}{0}{0+\dowOffSet}{0.5}{0.5}{0.25}{0.25}{0}
\draw[thick,red] (0.25+1,0.5+\dowOffSet)--(0.25+1,0.75+\dowOffSet);
\draw[thick,-] (0+1,0.25+\dowOffSet)--(-0.35+1,0.25+\dowOffSet);
%\draw[thick,-<] (0.5+1,0.25+\dowOffSet)--(0.5+0.25+1,0.25+\dowOffSet);
\draw[thick,-] (0.5+1,0.25+\dowOffSet)--(0.5+0.35+1,0.25+\dowOffSet);
\MPSxOnexD{A}{+1}{0+\dowOffSet}{0.5}{0.5}{0.25}{0.25}{0}
\MPSxOnexDColor{h}{0}{0.75+\dowOffSet}{1.5}{0.5}{0}{0}{0}{myred}
\draw[thick,red] (0.25,0.5+0.75+\dowOffSet)--(0.25,0.75+0.75+\dowOffSet);
\draw[thick,red] (0.25+1,0.5+0.75+\dowOffSet)--(0.25+1,0.75+0.75+\dowOffSet);
\node[right] at (2,0.24) {$=$};
\draw[thick,red] (0.25+3,0.5)--(0.25+3,0.75);
\draw[thick,-] (0+3,0.25)--(-0.35+3,0.25);
%\draw[thick,->] (0+3,0.25)--(-0.25+3,0.25);
%\draw[thick,-<] (0.5+3,0.25)--(0.5+0.25+3,0.25);
\draw[thick,-] (0.5+3,0.25)--(0.5+0.35+3,0.25);
\MPSxOnexDBlue{\scalebox{0.9}{$E$}}{+3}{0}{0.55}{0.5}{0.25}{0.25}{0}
\draw[thick,red] (0.25+4,0.5)--(0.25+4,0.75);
\draw[thick,-] (0+4,0.25)--(-0.35+4,0.25);
%\draw[thick,-<] (0.5+4,0.25)--(0.5+0.25+4,0.25);
\draw[thick,-] (0.5+4,0.25)--(0.5+0.35+4,0.25);
\MPSxOnexD{A}{+4}{0}{0.5}{0.5}{0.25}{0.25}{0}
\node[right] at (5,0.25) {$-$};
\pgfmathsetmacro{\x}{6}
\draw[thick,red] (0.25+\x,0.5)--(0.25+\x,0.75);
%\draw[thick,->] (0+\x,0.25)--(-0.25+\x,0.25);
\draw[thick,-] (0+\x,0.25)--(-0.35+\x,0.25);
%\draw[thick,-<] (0.5+\x,0.25)--(0.5+0.25+\x,0.25);
\draw[thick,-] (0.5+\x,0.25)--(0.5+0.35+\x,0.25);
\MPSxOnexD{A}{\x}{0}{0.5}{0.5}{0.25}{0.25}{0}
\pgfmathsetmacro{\x}{7}
\draw[thick,red] (0.25+\x,0.5)--(0.25+\x,0.75);
\draw[thick,-] (0+\x,0.25)--(-0.35+\x,0.25);
%\draw[thick,-<] (0.5+\x,0.25)--(0.5+0.25+\x,0.25);
\draw[thick,-] (0.5+\x,0.25)--(0.5+0.35+\x,0.25);
\MPSxOnexDBlue{\scalebox{0.9}{$E$}}{\x}{0}{0.55}{0.5}{0.25}{0.25}{0} 
    \end{tikzpicture}    
\end{aligned}
\end{equation*}
%We refer to the right hand side in eqn \ref{dksndjsnfjdsnfjdnfjdnfj} as the error term. 
When $H$ acts on $\ket{\Psi}$ the error terms cancel by translational invariance and one obtains
\begin{align}
H\ket{\Psi}=0\ .
\end{align}
Our ansatz by construction gives a vanishing energy eigenvalue and reduces the problem of constructing exact MPS eigenstates to finding representations of the quadratic algebra (\ref{dksndjsnfjdsnfjdnfjdnfj}). For a given Hamiltonian the tensors $A$ and $E$ provide us with $2d\chi^2$ free parameters, which are subject to $d^2\chi^2$ relations, so finding solutions is a non-trivial problem. We have obtained a number of solutions and now present representative cases. 
%%%%%%%%%%%%%%%%%%%%%%%%%%%%%%%%%%%%%%%%%%%%%%%%%%%%%%%%%
\subsection{Model I: degenerate multiplet of scars}
%%%%%%%%%%%%%%%%%%%%%%%%%%%%%%%%%%%%%%%%%%%%%%%%%%%%%%%%%
Our first example is a spin-S chain with a Hamiltonian that combines a generalized Rydberg term with a Dzyaloshinskii–Moriya (DM) interaction
%\begin{equation}\label{nfjdfnjdnfdjhh}
%H=\sum_{j=1}^N (1-\sigma^z_j)(1-\sigma^z_{j+1})+
%\vec{D}\cdot(\boldsymbol{\sigma}_j\times \boldsymbol{\sigma}_{j+1})\ ,
%\end{equation}
\begin{equation}
\label{nfjdfnjdnfdjhh}
H=\sum_{j=1}^N \bigg(1-\frac{S^z_j}{S}\bigg)\bigg(1-\frac{S^z_{j+1}}{S}\bigg)+
4\vec{D}\cdot(\boldsymbol{S}_j\times \boldsymbol{S}_{j+1})\ ,
\end{equation}
%where $\vec{D} = (D_x, D_y, D_z)$ and $\boldsymbol{\sigma}=(\sigma^{x},\sigma^{y},\sigma^{z})$ denotes the vector of Pauli matrices. 
where $\vec{D} = (D_x, D_y, D_z)$, $\boldsymbol{S}=(S^{x},S^{y},S^{z})$ denotes the vector of spin-S operators and we take $N$ to be even.
We note that the Hamiltonian (\ref{nfjdfnjdnfdjhh}) has a continuous spin-rotational symmetry only in the special case when $\vec{D}$ points along the z-direction.
We start by considering the simplest case $S=1/2$. A solution to eqn \eqref{dksndjsnfjdsnfjdnfjdnfj} for the Hamiltonian density of \eqref{nfjdfnjdnfdjhh} is given by
\begin{align}\label{fjsnjsnfjsfnj}
A=\begin{pmatrix}
 \ket{\uparrow}+\frac{\ri}{a} \ket{\downarrow}& \frac{1}{a}\ket{\downarrow}\\
\frac{1}{a}\ket{\downarrow}& \frac{b}{a}\ket{\uparrow}- \frac{\ri}{a}\ket{\downarrow}
\end{pmatrix}\,, \
\end{align}
where $a$ is a free complex parameter and 
$b=\frac{a}{1-a\Delta}$, $\Delta=\frac{D_y-\ri D_x}{D_z}$.
%\begin{align}\label{fkjdnfjdnfjdnfjfndjfn}
%b=\frac{a}{1-a\Delta}\,,\qquad \Delta=\frac{D_y-\ri D_x}{D_z}\,.
%\end{align}
The elements of the tensor $E$ are given by 
\begin{align}
E^1_{11} &= \frac{\ri \alpha}{a}+a\beta+\frac{2\ri D_zb}{a},\
E_{11}^2=\ri \beta+\alpha^*+\frac{2b\alpha^*}{a}\ ,\nonumber\\
E_{22}^1&= -\frac{\ri\alpha}{a}+b\beta+2\ri D_z,\ 
E_{22}^2=-\ri\beta+2\alpha^*+\frac{b\alpha^*}{a},\nonumber\\
E_{12}^1&=E_{21}^1=\frac{\alpha}{a}\ ,\ E_{12}^2=E_{21}^2=\beta\ ,
\end{align}
where 
\begin{equation}
\alpha=-(D_y+\ri D_x)\ ,\qquad \beta=\frac{D_x^2+D_y^2}{D_x}.
\end{equation}
The state $|\Psi\rangle$ exhibits a finite correlation length. In particular connected two-point functions decay exponentially at large separations $\langle\Psi|\sigma^{\alpha}_1\sigma^{\alpha}_{r+1}|\Psi\rangle_c\propto e^{-r/\xi}$. An explicit expression for $\xi$ can be readily obtained by standard MPS methods.
%%%%%%%%%%%%%%%%%%%%%%%%%%%%%%%%%%%
\subsubsection{Degenerate multiplet}
%%%%%%%%%%%%%%%%%%%%%%%%%%%%%%%%%%%
We observe numerically that the zero energy eigenvalue of (\ref{nfjdfnjdnfdjhh}) is $N/2+2$-fold degenerate. Out of these states $N/2+1$ have zero momentum and interestingly all can be obtained from our MPS as follows. We first note that the parameter $b$ can be expanded in a power series in $a^{-1}$ around $a=\infty$ as
$b=-\frac{1}{\Delta}\sum_{n=0}^\infty(a\Delta)^{-n}\,.$
This in turn provides us with a series expansion of the state $|\Psi\rangle$
\be
|\Psi\rangle=\sum_{n\geq 0}a^{-n}|v_n\rangle .
\label{expansion}
\ee
As $a$ is a free parameter that does not enter the expression of $H$ the $|v_n\rangle$'s must be zero energy eigenstates and lie within the Rydberg subspace, which forbids neighbouring up spins. Writing 
\begin{align}
A&=A_1+\frac{b}{a}A_2+\frac{1}{a}A_0\ ,\nonumber\\
A_1&=e^{11}\ket{\uparrow}\ ,\quad A_2=e^{22}\ket{\uparrow}\ ,\quad A_0=\sigma\ket{\downarrow}
\end{align}
with $\sigma=\sigma^x+\ri\sigma^z$
we see that $A_1A_2=0=A_2A_1=A_0A_0$. These restrict the non-vanishing terms in the expansion (\ref{expansion}) to those where $A_2$ is never adjacent to $A_1$ and $A_0$ is never followed by $A_0$. We conclude that 
\begin{enumerate}
\item{}The leading term in the expansion is the fully polarized ferromagnetic state in the z-direction $\ket{v_0}=\ket{\Uparrow}$.
\item{} For $1\leq n\leq N/2$ $\ket{v_n}$ contains contributions with at most $n$ overturned spins. This implies that $\ket{v_n}$ is linearly independent from $\{\ket{v_0},\dots,\ket{v_{n-1}}\}$, and the expansion (\ref{expansion}) generates the entire degenerate multiplet of $N/2+1$ zero momentum, zero energy eigenstates of $H$. 
\end{enumerate}
The structure of the degenerate multiplet is somewhat reminiscent of the one found in spin-1/2 XYZ chain \cite{BAXTER19731,Granovskii1985PeriodicSpinChain,PismaZhETF.41.312,bhowmick2025granovskii}. Explicit expressions for $|v_{n\leq 5}\rangle$ can be worked out in a straightforward fashion. It is useful to define generalized spin-lowering operators 
\begin{align}
{\cal S}^-&=\sum_j e^{21}_{j+1}\ ,\nn
{\cal S}_{2,n}^-&=\sum_j e^{21}_je^{21}_{j+n+1}\ ,\nonumber\\
%e^{11}_je^{21}_{j+1}e^{11}_{j+2}e^{21}_{j+3}e^{11}_{j+4}\ ,\nonumber\\
%{\cal S}_{2,2}^-&=\sum_j
%e^{11}_je^{21}_{j+1}e^{11}_{j+2}e^{11}_{j+3}e^{21}_{j+4}e^{11}_{j+5}\ .
{\cal S}_{3,1}^-&=\sum_j e^{21}_{j-2}e^{21}_{j}e^{21}_{j+2}\ .
\end{align}
A straightforward calculation then gives
\begin{align}
|v_1\rangle&\propto
  {\cal S}^-\ket{\Uparrow}\ ,\nonumber\\
|v_2\rangle&\propto\mathcal{P}_R({\cal S}^-)^2\ket{\Uparrow}\ ,\nonumber\\
|v_3\rangle&\propto\mathcal{P}_R\Big[\frac{\ri}{3!}({\cal S}^-)^3\ket{\Uparrow}+\frac{1}{\Delta}{\cal S}^-_{2,1}\ket{\Uparrow}\Big] ,\nn
|v_4\rangle&\propto\mathcal{P}_R\Big[\frac{1}{4!}({\cal S}^-)^4\ket{\Uparrow}
-\frac{\ri}{\Delta}\mathcal{S}^- \mathcal{S}_{2,1}\ket{\Uparrow}\nn
&\qquad\quad+\frac{1}{\Delta^2}({\cal S}^-_{2,2}-{\cal
  S}^-_{2,1})\ket{\Uparrow}\Big] ,\nn
|v_5\rangle&\propto\mathcal{P}_R\Big[\frac{\ri}{5!}({\cal S}^-)^5\ket{\Uparrow}+\frac{1}{2\Delta}(\mathcal{S}^-)^2\mathcal{S}_{2,1}\nonumber\\
&\qquad\quad+\frac{\ri}{\Delta^2}(\mathcal{S}^-(\mathcal{S}_{2,2}-\mathcal{S}_{2,1})-{\cal S}^-_{3,1})\ket{\Uparrow}\nonumber\\
&\qquad\quad-\frac{1}{\Delta^3}({\cal S}^-_{2,1}-2{\cal S}^-_{2,2}+{\cal S}^-_{2,3})\ket{\Uparrow}\Big],
\end{align}
where $\mathcal{P}_R$ is the projection operator on the Rydberg subspace.

We note that in Refs~\cite{kunimi2024proposal,kunimi2025systematic} a scar multiplet in the special case $D_x=D_z=0$  was constructed by a different method \footnote{The generalized Rydberg term becomes a conserved quantity in this case and does not alter the scar structure.}.
%%%%%%%%%%%%%%%%%%%%%%%%%%%%%%%%%%%%%%%%%%
\subsubsection{Open Boundary Conditions}
%%%%%%%%%%%%%%%%%%%%%%%%%%%%%%%%%%%%%%%%%%
So far we have focused on periodic boundary conditions. We now prove the existence of an exact MPS scar for 
the Hamiltonian (\ref{nfjdfnjdnfdjhh}) on an open chain with $N$ sites and certain boundary magnetic fields of the form
$\boldsymbol{\rm h}_1 \cdot \boldsymbol{\sigma}_1+\boldsymbol{\rm h}_N \cdot \boldsymbol{\sigma}_N$. To that end we consider an MPS of the form $|\Psi\rangle=BA\dots AC$, where $B$ and $C$ are tensors that only act in the auxiliary matrix space
\begin{equation}
\begin{aligned}
&\begin{tikzpicture}
\pgfmathsetmacro{\x}{1.75}
\node at (-1.75,0.25) {$\ket{\Psi}=$};
\draw[thick,red] (0.25,0.5)--(0.25,0.75);
\draw[thick,-] (0,0.25)--(-0.25,0.25);
\draw[thick,-] (0,0.25)--(-0.35,0.25);
\draw[thick,-] (0.5,0.25)--(0.5+0.25,0.25);
\draw[thick,-] (0.5,0.25)--(0.5+0.35,0.25);
\MPSxOnexD{A}{0}{0}{0.5}{0.5}{0.25}{0.25}{0}
\MPSxOnexD{B}{-1}{0}{0.5}{0.5}{0}{0.25}{0}
\node at (1.1,0.25) {$\tightldots$};
\draw[thick,red] (0.25+\x,0.5)--(0.25+\x,0.75);
\draw[thick,-] (0+\x,0.25)--(-0.25+\x,0.25);
\draw[thick,-] (0+\x,0.25)--(-0.35+\x,0.25);
\draw[thick,-] (0.5+\x,0.25)--(0.5+0.25+\x,0.25);
\draw[thick,-] (0.5+\x,0.25)--(0.5+0.35+\x,0.25);
\MPSxOnexD{A}{0+\x}{0}{0.5}{0.5}{0.25}{0.25}{0}
\MPSxOnexD{C}{1+\x}{0}{0.5}{0.5}{0.25}{0}{0}
\end{tikzpicture}  
\end{aligned}
\end{equation}
An exact energy eigenstate is obtained by same tensor $A$ as in (\ref{fjsnjsnfjsfnj}) with parameter $a$ fixed as $a=2/\Delta$, boundary vectors $B_1=C_1=1$, $B_2=-C_2=-\ri$ and then fine-tuning the boundary fields
\begin{align}
%\tilde{h}^0_N &= -\frac{
%D_y^3 - 3 D_y D_z^2 + D_y^2 \tilde{h}^x_N + D_x^2 (D_y + \tilde{h}^x_N) + %D_z^2 \tilde{h}^x_N
%}{
%2 D_x D_z
%}, \\[6pt]
{\rm h}^y_N &=\beta+\frac{D_y{\rm h^x_N}}{D_x} ,\
%\frac{ D^2_x +D_y (D_y + \tilde{h}^x_N)}{D_x}, \
{\rm h}^z_N = -\frac{
(D_x^2 + D_y^2 - D_z^2)(D_y + {\rm h}^x_N)
}{
2 D_x D_z
}, \nonumber\\
%\tilde{h}^0_1 &= -\frac{2 D_y D_z + D_x \tilde{h}^z_1}{D_x}, \qquad 
{\rm h}^x_1 &= -D_y, \qquad {\rm h}^y_1 = D_x.
%\\
%a &= \frac{2 D_y D_z}{D_x^2 + D_y^2}
%+ \ri \frac{2 D_x D_z}{D_x^2 + D_y^2}\,,\\
\end{align}
The components ${\rm h}^z_{1}$ and ${\rm h}^x_N$ remain arbitrary and the non-degenerate energy eigenvalue is
\begin{equation}
E=\frac{D_yD_z}{2D_x}(1+|\Delta|^2)+{\rm h}^z_1+\frac{{\rm h}^x_N|\vec{D}|^2}{2D_xD_z}.
\end{equation}

%%%%%%%%%%%%%%%%%%%%%%
\section{Higher spins \texorpdfstring{$S \geq 1$}{S >= 1}}
%%%%%%%%%%%%%%%%%%%%%%%
We now turn to higher spins $S\geq 1$. Numerical results for short chains suggest that the corresponding Hamiltonian again has an extensive number $N/2+2S$ of translation-invariant zero-energy eigenstates that only involve spin configurations that fulfil the generalized Rydberg constraint, i.e. neighbouring spin excitations are suppressed due to the interaction penalty. We have verified for $S\leq 2$ that our construction provides an exact MPS eigenstate in this subspace if we choose
\begin{align}
A=\begin{pmatrix}
 \ket{S}+\frac{\ri}{a} \ket{S-1}& \frac{1}{a}\ket{S-1}\\
\frac{1}{a}\ket{S-1}& \frac{b}{a}\ket{S}- \frac{\ri}{a}\ket{S-1}
\end{pmatrix}\ ,
\end{align} 
where $|m\rangle$ denotes the $S^z$ eigenstate with eigenvalue $m$ and impose $b=\frac{a}{1-a\Delta_S}$ with $\Delta_S= \frac{\Delta}{\sqrt{2S}}$.
We omit the explicit expressions for the $E$ tensors, as they become increasingly cumbersome with increasing $S$. 
Expanding the MPS in a power series in $1/a$ around zero provides us with $N/2+1$ linearly independent states in the degenerate multiplet in a way analogous to the $S=1/2$ case. It would be interesting to prove that our results extend to $S>2$.

%%%%%%%%%%%%%%%%%%%%%%%%%%%%%%%%%%%%%%%
\subsection{Model II: isolated scar.}
%%%%%%%%%%%%%%%%%%%%%%%%%%%%%%%%%%%%%%%
Our second example is a spin-1 chain with Hamiltonian 
\begin{align}
H&=\sum_{j=1}^N\Big[\tfrac{J_y-J_z}{2}(\mathbf{S}_j \cdot \mathbf{S}_{j+1}-(\mathbf{S}_j \cdot \mathbf{S}_{j+1})^2)
+2h_y S^y_j\nonumber\\
&+2J_z[(S^x_j)^2+(S^z_j)^2]+2J_y(S^y_j)^2-J_y-3J_z\Big],
\end{align}
where $J_{y,z}$ and $h_y$ are free parameters. Our construction (\ref{dksndjsnfjdsnfjdnfjdnfj}) provides us with a zero energy MPS eigenstate with tensors $A$ and $E$ given by
\begin{equation}\label{dcjcdncjdncjdncjdncj}
A=\begin{pmatrix}
\sqrt{2}\ket{-1}& \ket{0}\\
\ket{0}&\sqrt{2}\ket{1}
\end{pmatrix},\ 
E=\ri h_y\begin{pmatrix}
0&-|+\rangle\\
|+\rangle&0
\end{pmatrix},
\end{equation}
where $|+\rangle=\frac{1}{\sqrt{2}}(|1\rangle+|-1\rangle)$.
By exact diagonalisation for small sizes $N$ we find that the eigenspace of eigenvalue $0$ is non-degenerate. Standard numerical MPS calculations indicate that the state exhibits a finite non-vanishing correlation length $\xi=\log(3)$ for connected correlators.
%%%%%%%%%%%%%%%%%%%%%%%%%%%%%%%%%%%%%%%%%%%%%%%%%%%%%%%%
\section{The method on the square lattice.} 
%%%%%%%%%%%%%%%%%%%%%%%%%%%%%%%%%%%%%%%%%%%%%%%%%
We now turn to Hamiltonians of the form
$H=\sum_{\braket{i,j}} \, h_{i,j}$, where $\braket{i,j}$ denotes nearest neighbours on the square lattice. We construct translationally invariant zero-energy MPS eigenstates (\ref{MPSform}) described by the local tensor $[A^s]^{b_x\,b_y}_{a_x\,a_y}$, where $s$ is the physical index:
\begin{align}
\begin{tikzpicture}[baseline={(0.25,0,0.5)}]
\node at (-2.,0,0.25) {$[A^s]^{b_x\,b_y}_{a_x\,a_y}=$};
\PEPS{0}{0}{0}{0.75}{0.25}{0.25}
\node[above] at (0.35,0.3,0.375) {$s$}; 
\node[right] at (-0.75,0,0.4) {$a_x$}; 
\node[right] at (1.0,0,0.4) {$b_x$}; 
\node[right] at (0.2,0,1.5) {$a_y$}; 
\node[right] at (0.1,-0.05,-0.7) {$b_y$}; 
\end{tikzpicture}
\end{align}
The MPS can be represented graphically as 
\begin{align}\label{sdjsnjsndjsdn}
\begin{tikzpicture}[baseline={(0,0,2.25)}]
\node at (-1,0,2.25) {$\ket{\Psi}=$};
\foreach \varX in {0,1,2,3,4} {
\foreach \varZ in {0,1,2,3,4} {
    \PEPS{0.75*\varX}{0}{\varZ}{0.5}{0.25}{0.25}
    }
}
\end{tikzpicture}
\end{align}
where indices are contracted along nearest-neighbour bonds in the $x$ and $y$ directions and we impose PBC. 
In this geometry our ansatz (\ref{gnkdnrnr}) can be cast in the form
\begin{equation}
\begin{aligned}
\begin{tikzpicture}[baseline={(0,-1,0)}]
\PEPSY{0.75}{0}{0}{0.5}{0.2}{0.25}{1.2}
\PEPSY{0.75}{0}{1}{0.5}{0.2}{0.25}{1.2}
\node at (2,-0.3) {$=$};
\PEPSC{3+1}{0}{0}{0.5}{0.2}{0.25}{myblue}
\PEPSC{3+1}{0}{1}{0.5}{0.2}{0.25}{myolive}
\node at (3.85+1,-0.3) {$+$};
\PEPSC{5+1}{0}{0}{0.5}{0.2}{0.25}{myolive}
\PEPSC{5+1}{0}{1}{0.5}{0.2}{0.25}{mypurple}

\def\x{0.70}
\def\y{0.1}
\def\z{0.85}
\draw[thick, rounded corners=1pt, fill=myred]
  (\x,\y,\z) -- (\x,\y+0.5,\z) -- (\x,\y+0.5,\z-1.7) -- (\x,\y,\z-1.65) -- cycle;

%%%%%%%%%%%%%%%%%%%%%%%%%%%%%%%%%%%%%%%%%%%%%

\PEPSY{0}{-2}{0}{0.5}{0.2}{0.25}{1.}
\PEPSY{0.75}{-2}{0}{0.5}{0.2}{0.25}{1.}
\node at (2,-2.1) {$=$};

\PEPSC{3}{-2}{0}{0.5}{0.2}{0.25}{myyellow}
\PEPS{3.75}{-2}{0}{0.5}{0.2}{0.25}{1.2}
\node at (4.625,-2.1) {$+$};
\PEPS{3+2.25}{-2}{0}{0.5}{0.2}{0.25}{1.2}
\PEPSC{3.75+2.25}{-2}{0}{0.5}{0.2}{0.25}{myorange}

\def\x{0}
\def\y{-1.85}
\def\z{0}
\draw[thick, rounded corners=1pt, fill=myred]
  (\x,\y,\z) -- (\x,\y+0.5,\z) -- (\x+1.05,\y+0.5) -- (\x+1.05,\y) -- cycle;

\end{tikzpicture}
\end{aligned}
\label{DEHP2D}
\end{equation}
where the blue, purple, yellow and orange coloured squares represent four distinct tensors $E^{(1)}$, $E^{(2)}$, $E^{(3)}$ and $ E^{(4)}$. The local cancellation condition 
\begin{align}\label{fkjndjfndjfndjfndj}
E^{(1)}+E^{(2)}+E^{(3)}+E^{(4)}=0
\end{align}
ensures that that $|\Psi\rangle$ is an exact zero-energy eigenstate.
%%%%%%%%%%%%%%%%%%%%%%%%%%%%%%%%%%%%%%%%%%%%%%%%%%
\subsection{Spin-2 model on the square lattice}
%%%%%%%%%%%%%%%%%%%%%%%%%%%%%%%%%%%%%%%%%%%%%%%%%%
We now fix the spin $S=2$ and bond dimension $\chi=2$. We denote the eigenstates of $S^z$ on a single site by 
$\{\ket{2}, \ket{1}, \ket{0}, \ket{-1}, \ket{-2}\}$ and use notations where $\bar{a}=-a$. We take the Hamiltonian density to be 
\begin{equation}
h_{j,k}=\sum_{a=0}^4\lambda_a\bigl(P_{j,k}^{(a)}+P_{j,k}^{(\bar{a})}\bigr)+h^z\, (S^z_j+S^z_{k})
\label{H2D}
\end{equation}
where $P^{(a)}_{j,k}=|a\rangle_j\ {}_k\langle a|$ are projection operators on $S^z_j+S^z_k$-eigenstates on the link $\langle j,k\rangle$ with eigenvalues $a$
\begin{align}
\lvert 4\rangle &= \lvert 2,2 \rangle\,, \quad
\lvert 3 \rangle = \lvert 1,2 \rangle - \lvert 2,1 \rangle\ \nonumber \\
\lvert 2 \rangle &=\frac{b}{a^2}\,\lvert 1,1 \rangle- \lvert 2,0 \rangle  - \lvert 0,2 \rangle \nonumber \\
\lvert 1 \rangle &= \lvert 2,\bar{1} \rangle - b\,\lvert 1,0 \rangle + b\,\lvert 0,1 \rangle - \lvert \bar{1},2 \rangle  \nonumber\\
\lvert 0 \rangle &= \frac{1}{b^2}(\lvert 2,\bar{2} \rangle +\lvert \bar{2},2 \rangle)
- \frac{1}{a^2}(\lvert 1,\bar{1} \rangle+\lvert \bar{1},1 \rangle ) + \lvert 0,0 \rangle \nonumber
\end{align}
Applying our construction gives an exact zero-energy MPS eigenstate of this Hamiltonian if we set  $[A^{a}]^{\alpha\beta}_{\gamma\delta} =[A^{\bar{a}}]_{\alpha\beta}^{\gamma\delta}$ and
%\begin{equation}
%\begin{aligned}
%[A^{0}]^{11}_{11} &= 1      &\quad [A^{\bar{1}}]^{11}_{12} &= a    &\quad [A^{1}]^{12}_{11} &=  a \\
%[A^{0}]^{12}_{12} &= 1           &\quad [A^{\bar{1}}]^{11}_{21} &= a           &\quad [A^{\bar{2}}]^{11}_{22} &= b \\
%[A^{0}]^{12}_{21} &= 1           &\quad [A^{\bar{1}}]^{12}_{22} &= a    &\quad [A^{1}]^{21}_{11} &= a \\
%[A^{0}]^{21}_{12} &= 1           &\quad [A^{2}]^{22}_{11}  &= b           &\quad [A^{1}]^{22}_{12} &= a \\
%[A^{0}]^{21}_{21} &= 1           &\quad [A^{\bar{1}}]^{21}_{22} &=  a    &\quad [A^{1}]^{22}_{21} &= a \\
%[A^{0}]^{22}_{22} &= 1
%\end{aligned}
%\end{equation}
\begin{align}
[A^{2}]^{22}_{11}  = b ,\ 
[A^{1}]^{22}_{\alpha\hat{\alpha}} =[A^{1}]^{\alpha\hat{\alpha}}_{11} = a,\
[A^{0}]^{\alpha\beta}_{\beta\alpha} =[A^{0}]^{\alpha\beta}_{\alpha\beta}  = 1,  \label{A2D} 
\end{align}
where $\hat{\alpha}=3-\alpha$. The A-tensors fulfil the local algebra \eqref{DEHP2D}, where the non-zero elements of the $E$ tensors are given by
\begin{align}
[(E^{(1)})^{{1}}]^{12}_{11}&=[(E^{(1)})^{{1}}]^{21}_{11}=[(E^{(1)})^{{1}}]^{22}_{12}=ah^z\ ,\nonumber\\
[(E^{(1)})^{\bar{1}}]^{11}_{21}&=[(E^{(1)})^{\bar{1}}]^{12}_{22}=[(E^{(1)})^{\bar{1}}]^{21}_{22}=3ah^z\ ,\nonumber\\
[(E^{(1)})^{0}]^{21}_{21}&=[(E^{(1)})^{0}]^{22}_{22}=[(E^{(1)})^{0}]^{12}_{21}=4h^z\ ,\nonumber\\
[(E^{(1)})^{\bar{2}}]^{11}_{22}&=[(E^{(1)})^{2}]^{22}_{11}=2bh^z\ ,\nonumber\\
[(E^{(1)})^{\bar{1}}]^{11}_{12}&=-ah^z\ ,\ 
[(E^{(1)})^{{1}}]^{22}_{21}=5ah^z\ .
\end{align}
\begin{align}
[(E^{(2)})^{{a}}]^{\alpha\alpha'}_{\beta\beta'}&=-[(E^{(1)})^{\bar{a}}]_{\alpha\alpha'}^{\beta\beta'}\ ,\nonumber\\
[(E^{(3)})^{{a}}]^{\alpha\alpha'}_{\beta\beta'}&=[(E^{(1)})^{{a}}]^{\alpha\alpha'}_{\beta'\beta}
\end{align}
The MPS (\ref{sdjsnjsndjsdn}) with the same tensor $A$ (\ref{A2D}) was previously shown to be a frustration-free ground state of the Hamiltonian (\ref{H2D} for vanishing magnetic field $h^z=0$ \cite{Niggemann2000}. A non-zero field $h^z\neq 0$  breaks the spin-flip symmetry of $H$ and $|\Psi\rangle$ turns into a quantum many-body scar, but loses the frustration-free property.
The physical properties of $|\Psi\rangle$ and in particular its correlation length have been discussed in \cite{Niggemann2000}. 
%%%%%%%%%%%%%%%%%%%%%%%%%%%%%%%%%%%%%%%%%%%%%%%%%%
\subsection{Spin-1 XYZ model with DM interaction}
%%%%%%%%%%%%%%%%%%%%%%%%%%%%%%%%%%%%%%%%%%%%%%%%%%
Our final example is the $S=1$ XYZ model with DM interaction and a magnetic field in the $z$-direction with even numbers of sites in both directions of the square lattice
\begin{align}
h_{k,\ell}=&\,\mathcal{C}\, \mathbbm{1}+J_{x}\, S^x_kS^x_\ell+J_{y}\, S^y_kS^y_\ell+J_{z}\, S^z_k S^z_\ell\\
&+D_{xy} (S^x_kS^y_\ell-S^y_kS^x_\ell)+h_z(S^z_k+S^z_\ell)
\end{align}
where $h_z=\pm\sqrt{(J_x + J_z) (J_y + J_z)})$ and $\mathcal{C}=\,J_x+J_y+J_z$. We assume from hereon that the $J_\alpha$ are chosen such that $h_z$ is real. An exact zero-energy eigenstate of MPS form is obtained by the choice
\begin{align}
\label{mdfdnfjdnfjdnfj}
[A]^{b_x\,b_y}_{a_x\,a_y}=&\begin{pmatrix}
[B_+]^{b_x}_{a_x}&[B_-]^{b_x}_{a_x}\\
[B_-]^{b_x}_{a_x}&-[B_+]^{b_x}_{a_x}
\end{pmatrix}^{b_y}_{a_y},\
B_\pm=\begin{pmatrix}
\ket{\pm}&\pm\ket{\mp}\\
\pm\ket{\mp}&-\ket{\pm}
\end{pmatrix},
\end{align}
where $\ket{+}=z_+\ket{1}+\ket{-1}$, $\ket{-}=\sqrt{-2z_+}\ket{0}$ with
$z_+=[(J_x+J_y+2J_z)-2h_z](J_x-J_y)^{-1}$.
The error terms are given by $E_1=-E_2=E_3=-E_4$ and $E_1$ has the same form as \eqref{mdfdnfjdnfjdnfj}  with the replacements $|+\rangle\rightarrow-2\ri z_+ D_{xy}\ket{+1}$ and
$\ket{-}\rightarrow-\ri  D_{xy} \sqrt{-2z_+}\ket{0}$. Setting $D_{xy}=0$ leads to a frustration-free structure and the MPS becomes one of two degenerate ground states. For $D_{xy}\neq 0$ $|\Psi\rangle$ ceases to be a ground state, but can be decomposed into two product states, which describe the GS manifold in the frustration-free case. As a result the MPS has a vanishing correlation length for connected correlations in the infinite volume limit. 
%We have also checked this fact employing a corner transfer matrix renormalisation group algorithm of the PEPSKit \cite{Brehmer_PEPSKit_2025}.
%%%%%%%%%%%%%%%%%%%%%%%%
\section{Discussion}
%%%%%%%%%%%%%%%%%%%%%%%%
We presented a method for constructing exact matrix product eigenstates of quantum many-body Hamiltonians. 
Our approach makes an ansatz for the action of the Hamiltonian density that goes beyond the frustration-free setting by allowing for the generation a local "error" that telescopes to zero globally. Our framework can be viewed as a generalization of both the DHEP matrix product ansatz \cite{DerEvaHakPas1993}, and the method of Klümper and Zittartz  \cite{KluSchZit1993, NigKluZit1997,NigKluZit2000}, which are recovered by particular, simple choices of the error terms. Specifically, they corresponds to choosing the identity operator in the bond space and vanishing eror terms respectively.
We have applied our framework to both one and two-dimensional systems and constructed illustrative examples of models that harbour exact quantum many-body scars of exact MPS form. Given their low bond-dimension these scar states could be seen in the dynamics of local observables on present-day quantum computers.

During the writing of this manuscript we became aware of a recent preprint~\cite{FrankPaper}, which investigates a more general version of our Ansatz in the context of injective MPS.

\acknowledgments  {This work was supported by the EPSRC under grant EP/X030881/1. We thank Denis Bernard, Paul Fendley, Dominik Hahn, Andreas Kl\"umper, Patrick Penc, Werner Krauth, Frank Verstraete and Weronika Wiesiolek for helpful discussions.}

\bibliographystyle{apsrev4-2}
\bibliography{bib.bib}

%%%%%%%%%%%%%%%%%%%%%%%%%%%%%

%%%%%%%%%%%%%%%%%%%%%%%%%%%%%%

\end{document}